\documentclass[11pt]{article}
\pdfoutput=1
\def\be{\begin{equation}}
\def\ee{\end{equation}}
\def\bea{\begin{eqnarray}}
\def\eea{\end{eqnarray}}
\usepackage{graphicx}
\usepackage{amssymb,amsmath}
\usepackage{cite,color,url}
\usepackage[colorlinks=true
,urlcolor=blue
,anchorcolor=blue
,citecolor=blue
,filecolor=blue
,linkcolor=blue
,menucolor=blue
,linktocpage=true
,pdfproducer=medialab
,pdfa=true
]{hyperref}

\catcode`\@=11
\def\lsim{\mathrel{\mathpalette\@versim<}}
\def\gsim{\mathrel{\mathpalette\@versim>}}
\def\@versim#1#2{\vcenter{\offinterlineskip
\ialign{$\m@th#1\hfil##\hfil$\crcr#2\crcr\sim\crcr } }}
\catcode`\@=12
\catcode`@=12
\begin{document}
\thispagestyle{empty}
\begin{flushright}
\end{flushright}
\vspace{0.3in}
\begin{center}
{\Large \bf Slim SUSY \\} 
\vspace{1.0in}
{\bf Ernesto Arganda$^a$, J. Lorenzo Diaz-Cruz$^b$ and Alejandro Szynkman$^a$}
\vspace{0.2in} \\
{\sl $^a$ IFLP, CONICET - Dpto. de F\'{\i}sica, Universidad Nacional de La Plata, \\ 
C.C. 67, 1900 La Plata, Argentina} \\
{\sl $^b$ Facultad de Ciencias F\'{\i}sico-Matem\'aticas, \\
Benem\'erita Universidad Aut\'onoma de Puebla, Puebla, M\'exico }
\end{center}
\vspace{1.0in}
\begin{abstract}
The new SM-like Higgs boson discovered recently at the LHC, with mass $m_h \simeq$ 125 GeV,  as well as the direct LHC bounds on the mass of superpartners, which are entering into the TeV range, suggest that the minimal surviving supersymmetric extension of the SM (MSSM), should be characterized by a heavy SUSY-breaking scale. Several variants of the MSSM have been proposed to account for this result, which vary according to the accepted degree of fine-tuning. We propose an alternative scenario here, Slim SUSY,  which contains sfermions with multi-TeV masses and gauginos/higgsinos near the EW scale, but it includes the heavy MSSM Higgs bosons ($H^0$, $A^0$, $H^\pm$) near the EW scale too. We discuss first the formulation and constraints of the Slim SUSY scenario, and then identify distinctive heavy Higgs signals that could be searched at the LHC, within  scenarios with the minimal number of superpartners with masses near the EW scale.

\end{abstract}

\vspace*{30mm}
\noindent {\footnotesize E-mail:
{\tt \href{mailto:ernesto.arganda@fisica.unlp.edu.ar}{ernesto.arganda@fisica.unlp.edu.ar},
\href{mailto:jldiaz@fcfm.buap.mx}{jldiaz@fcfm.buap.mx},
\href{mailto:szynkman@fisica.unlp.edu.ar}{szynkman@fisica.unlp.edu.ar}}}

\newpage

\section{Introduction}
\label{intro}

The search for the Higgs boson and physics beyond the Standard Model (SM) were among 
the prime motivations to build the LHC. With the recent LHC discovery of a new particle with 
SM-like Higgs properties and a mass around
$m_{h_\text{SM}} \simeq$ 125 GeV~\cite{:2012gk,:2012gu}, the first mission seems to be accomplished.
The fact that the Higgs-like mass value agrees quite well with the range preferred
by the analysis of electroweak precision tests (EWPT)~\cite{Erler:2010wa},
could be seen as a confirmation of the SM. Further studies of the Higgs couplings
are required in order to confirm its SM nature~\cite{Espinosa:2012im,Giardino:2012ww},
or to find evidence of physics beyond the SM. In fact, the LHC has already provided important bounds
on the scale of new physics.
However, the failure of the LHC, so far, to find evidence of new particles
beyond the SM, has raised some premature concern.

Within the Minimal Supersymmetric Standard Model (MSSM),
which is the most popular realization of supersymmetry (SUSY)
at the electroweak (EW) scale,
the lightest Higgs boson mass satisfies the tree-level relation $m_{h^0} \leq$ $m_Z$,
while radiative corrections involving the top/stop system are needed
in order to bring $m_{h^0}$ above the LEP bound, $m_{h^0} >$ 115 GeV.
In fact, to make the MSSM light Higgs boson to reach a mass of 125 GeV,
one needs to include stop masses of order TeV and/or large values of tan$\beta$.
Similarly, the direct search for squarks and gluinos at the LHC is actually lifting 
their masses limits
to the multi-TeV range~\cite{MTDova-Silafae2012}.
Furthermore, the masses of all the MSSM particles must also agree with all bounds from collider
and low-energy frontiers,
and so far no effect has been detected 
that would require superpartners with masses below the TeV range,
with the possible exception of the anomalous magnetic moment of the muon.

These results suggest that SUSY is actually badly broken, though still softly,
and bring into question the original motivation to solve the hierarchy/naturalness problem,
as the resulting constraints are difficult to fulfill in the most constrained versions of the MSSM,
namely for the cMSSM or minimal SUGRA. Several avenues of reasoning have arisen in the SUSY community
to cope with this situation:

\begin{enumerate}
\item On one side there is the so-called phenomenological MSSM (pMSSM)~\cite{Djouadi:1998di,Djouadi:2002ze,Berger:2008cq},
which takes advantage of the large number of parameters that come with the MSSM.
Then, one looks for regions of the parameter space where the current bounds on
Higgs and SUSY are reproduced~\cite{pMSSM_works}.
This could be seen as a ``no compromise'' model, which will evolve as more data comes from the LHC.
\item From a point of view slightly different, Natural SUSY and its relatives~\cite{NaturalSUSY}
offer the possibility of keeping supersymmetry as a solution of   the hierarchy problem
without re-introducing fine-tuning, which was one of its main phenomenological motivations.
The paradigm of naturalness is actually in tension with the current direct SUSY bounds but 
it is still enduring~\cite{Espinosa:2012in,D'Agnolo:2012mj,Baer:2012cf}.
\item On the other hand, we have Split SUSY~\cite{SplitSUSY}, 
which falls under the enchantment of the landscape and the fine-tuning sirens.
Motivated by the present lack of explanation for the cosmological constant ($\Lambda$), one simply assumes that
nature accepts a couple of fine-tuning for $\Lambda$ and the Higgs mass.
Split SUSY models have been widely studied lately~\cite{SplitSUSY_works},
and assume that, except for the light SM-like Higgs boson ($h^0$), all scalars are in the multi-TeV range,
while gauginos and higgsinos would have lower masses and could be at the reach of the LHC.
Alternative models, inspired in the landscape philosophy of Split SUSY, have been also proposed, such as
Spread SUSY~\cite{SpreadSUSY} and High-Scale SUSY~\cite{High-ScaleSUSY}.
\end{enumerate}

However, this pattern of heavy sfermions has a positive side,
namely the possibility to solve the SUSY flavor and CP problems
by decoupling~\cite{ArkaniHamed:1997ab,DiazCruz:2005qz}.
And this reminds us that there are open problems within the SM, notably the CP and flavor problems,
whose solution may also leave its imprints in the parameters of the MSSM. But regarding the SUSY flavor problem,
we notice that the Higgs doublets are somehow harmless. In fact, one could have the full heavy Higgs spectrum ($H^0$, $A^0$ and $H^\pm$),
with masses near the EW scale without any phenomenological conflict~\cite{Gupta:2012fy}.
For instance, the approximate degeneracy between the heavy Higgs bosons facilitates the agreement with EWPT;
similar conclusion holds for the implications of the Yukawa couplings for low-energy flavor observables
and collider results~\cite{DiazCruz:2008ry}. Thus, one could imagine other reasons,
beyond the landscape and fine-tuning arguments, to have heavy sfermions in the MSSM. For instance, when one 
considers flavor symmetries, it happens usually that the quarks, leptons and their superpartners,
behave differently from the Higgs doublets, with the sfermions having flavor quantum numbers,
while the Higgs doublets are singlets. 
Consequently, they would have different behavior when the flavor symmetry is broken.

The aim of this paper is precisely to discuss the possible realization of scenarios
with Higgs masses near the EW scale (which here it means to be in the range $0.2-1$ TeV). In some sense we shall be
studying a type of two Higgs doublet model (2HDM) with MSSM parameters and additional states, 
including a dark matter (DM) candidate, the lightest supersymmetric particle (LSP), which we assume to be  
the neutralino ($\tilde \chi_1^0$), with MSSM parameters chosen such that 
$m_{\tilde \chi_1^0} =$ ${\cal O}$(100 GeV).

In our previous work~\cite{Arganda:2012qp}, we studied the effect of non-universal Higgs masses
within Split-inspired SUSY scenarios, focusing in heavy Higgs decays. However, after closely examining the 
defining hypothesis, we have learned now that our proposal goes beyond the Split SUSY models,
which are based on the landscape paradigm. In fact, ref.~\cite{Delgado:2005ek}, which clarifies the meaning 
of the fine-tuning associated with Split SUSY, also discussed briefly the possibility to have both Higgs 
doublets near the EW scale. This requires the imposition of a second fine-tuning, besides the one required
to have a light Higgs boson at the EW scale.
However, if the fine-tuning is a sign of exceptionality, we feel that using it twice would 
appear less motivated. Thus, we shall not associate the presence of the full Higgs 
spectrum of the MSSM near the EW scale with the philosophy of Split SUSY,
but rather as a sign that the MSSM is also part of a more fundamental theory, 
with an unknown sector that communicates SUSY breaking with the MSSM to 
make the sfermions quite heavy, while it leaves the Higgs doublets light 
enough to be searched at the LHC or future colliders.

The paper is organized as follows: in Section~\ref{SlimSUSY} we present
the Slim SUSY scenario and discuss its possible theoretical realizations and
its corresponding SUSY spectra.
Section~\ref{constraints} is devoted to the study of the constraints that
the current Higgs mass data and the strength of the SM-like Higgs signals
observed at the LHC impose over the proposed scenario.
We dedicate Section~\ref{results} to examine the decays and production of
heavy neutral Higgs bosons at the LHC in specific scenarios of Slim SUSY.
Finally, perspectives and conclusions are presented in Section~\ref{conclusions}.

\section{The Slim SUSY scenario}
\label{SlimSUSY}

The MSSM is considered as an attractive model not only because it realizes a new type of symmetry,
between bosons and fermions, that helps to solve the hierarchy problem,
but also because its building blocks (R-parity) allow for the presence of a DM candidate,
with the right mass and couplings to generate the measured relic density.
The model is also nice because it predicts gauge coupling unification at a scale that 
satisfies bounds on proton decay. The model also contains new sources of CP violation, 
which may allow to generate the right baryon asymmetry of the universe, while at the 
same time it should be free of the SUSY flavor and CP problems.

In order to account for all the above constraints and satisfy all the bounds on Higgs and SUSY at the LHC,
we shall define our Slim SUSY scenario, with the following assumptions:
\begin{enumerate}
\item It contains heavy sfermions of third generation (with $m =$ ${\cal O}$(TeV)), 
to account for the Higgs mass value ($m_{h_\text{SM}} \simeq$ 125 GeV).
\item Heavy masses of  ${\cal O}$(10$-$100 TeV) for the first and second generations of sfermions to solve the SUSY and CP flavor problems or at least to ameliorate them.
\item A neutralino sector with an LSP mass of ${\cal O}$(100 GeV), which is chosen as the DM 
candidate~\cite{Baer:2011ab}. Other possibilities, such as gravitino DM, could be acceptable too.
\item The full Higgs sector has masses near the EW scale.
\end{enumerate}

The main phenomenological motivation for this scenario is precisely the fact that this spectrum, with the complete MSSM Higgs sector having masses near the EW scale, has not been considered in detail before, and thus it should be explored at the LHC in order to fully test the possible realization of SUSY at the EW scale.

In order to provide a general definition of the parameter space of the Slim SUSY scenario,
we assume that all of the soft-masses of squarks and sleptons of the first and second generations
are given by only one parameter, $M_S$.
We also consider only a common soft mass for the third generation of sfermions,
$m_s$, which is defined as the boundary condition for the RGEs.
Therefore, the relevant MSSM parameters of Slim SUSY are the following:
\begin{itemize} 
\item $1 < \tan\beta < 60$,
\item 200 GeV $< m_{A^0} < 600$ GeV,
\item 0.1 TeV $< |M_1|$, $|M_2|$, $|\mu| < 3$ TeV,
\item  1 TeV $< M_3 < 3$ TeV,
\item $-10$ TeV $< A_t <$ 10 TeV,
\item 10 TeV $< M_S <$ 100 TeV,
\item 1 TeV $< m_s <$ 7.5 TeV,
\end{itemize}
where $\tan\beta$ is the ratio of the two Higgs vacuum expectation values,
$m_{A^0}$ is the mass of the pseudoscalar Higgs mass,
$A_t$ is the common trilinear coupling for the sfermions of the third generation
and $M_1$, $M_2$, $M_3$ and $\mu$ are the bino, wino, gluino and higgsino masses, respectively.
We notice that the Slim SUSY spectra are somehow similar to
the radiative natural SUSY ones~\cite{Baer:2012cf}. However, the sfermion sector
of the former is much heavier than the latter and it could be even heavier,
as LHC searches for SUSY are indicating, since we do not have to deal with
the constraints that naturalness imposes. Moreover, we would like to emphasize
that Slim SUSY is not a Split SUSY scenario either, since we do not decouple the heavy scalar states.

Although one expects that heavy sfermions would decouple from low energies,
there are RGE effects that could be important,
namely it is possible that $m_{A^0}$ acquires imaginary tree-level values
(meaning that the electroweak Higgs minimum is essentially unstable)
or the sfermion masses of the third generation may become tachyonic~\cite{tachyons}.
In our previous work~\cite{Arganda:2012qp}, we checked that these problems can be avoided
if we increase $m_{A^0}$ or decrease $m_s$,
and scenarios with values of $m_s$ below 8 TeV do not present this kind of difficulty.
However, it would be important to perform a thorough study of the constraints on our scenarios coming from
RGEs and correct EW symmetry breaking, which is out of the scope of this paper,
and we leave for future works.

\begin{figure}[t!]
\begin{center}
\begin{tabular}{c}
\includegraphics[width=100mm]{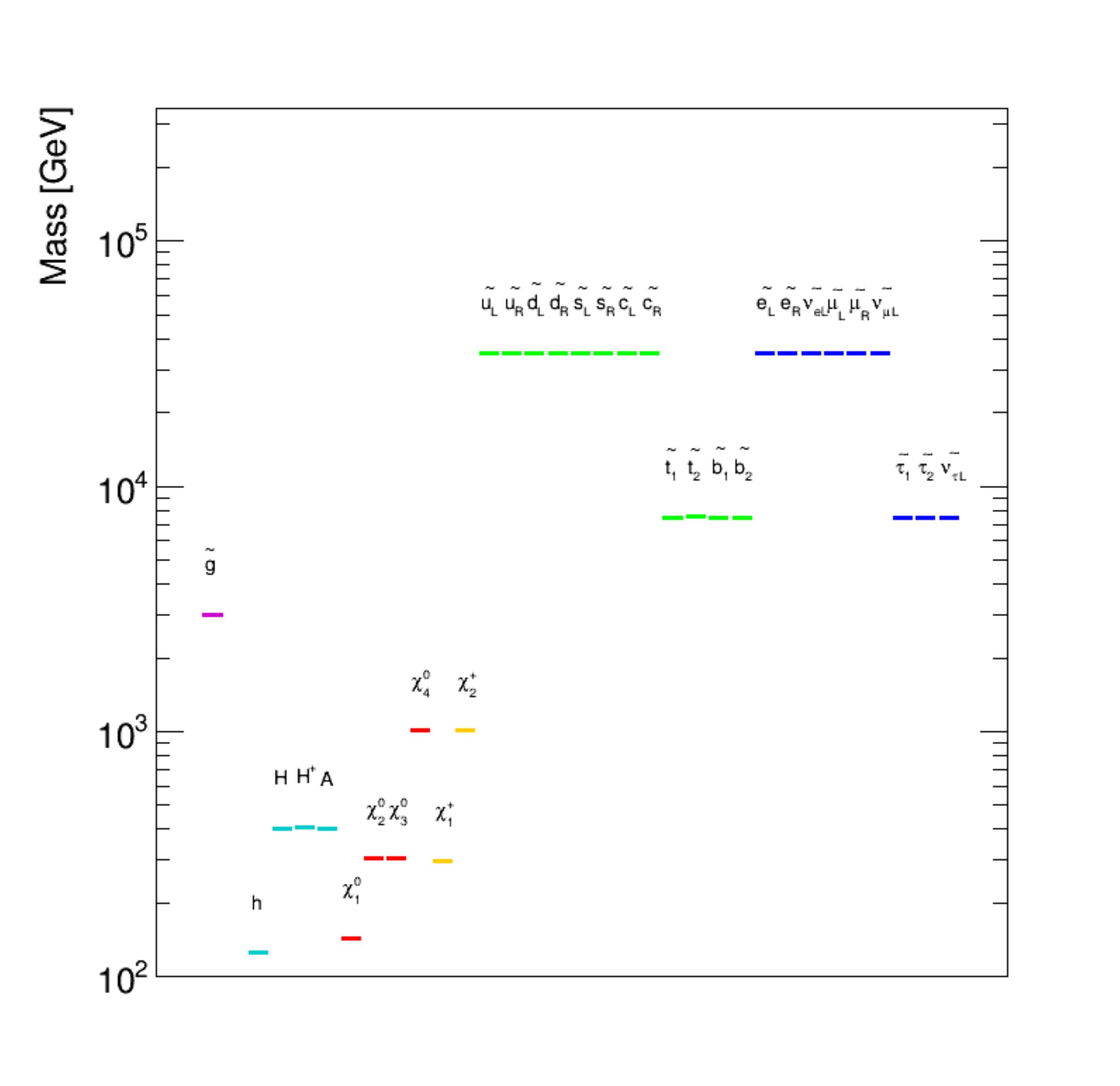}
\end{tabular}
\caption{Supersymmetric mass spectrum for $M_S =$ 35 TeV,
$m_s =$ 7.5 TeV, $m_{A^0} =$ 400 GeV, $\tan\beta =$ 7.5,
$A_t =$ 0, $M_1 =$ 150 GeV, $M_2 =$ 1 TeV,
$M_3 =$ 3 TeV and $\mu =$ 300 GeV.}\label{fig:mass_spectrum}
\end{center}
\end{figure}

Furthermore, to illustrate the type of supersymmetric spectrum
arisen within Slim SUSY, we have displayed in Figure~\ref{fig:mass_spectrum}
the full spectrum of squarks, sleptons, charginos, neutralinos and Higgs bosons,
for the following choice of parameters: $M_S =$ 35 TeV, $m_s =$ 7.5 TeV,
$m_{A^0} =$ 400 GeV, $\tan\beta =$ 7.5, $A_t =$ 0, $M_1 =$ 150 GeV, $M_2 =$ 1 TeV,
$M_3 =$ 3 TeV and $\mu =$ 300 GeV.
This choice represents a SUSY point with bino-like LSP (with large higgsino admixture)
and the only supersymmetric signals available for the current energies at the LHC
are the invisible decays of $H^0$ and $A^0$ Higgs bosons into
two LSP neutralinos. The rest of neutralinos and charginos are too heavy
to be produced through the decays of the heavy Higgs bosons, and
the gluino, sleptons and squarks are not reachable at the LHC.
We shall analyze in more detail the possible signals of this class of
SUSY spectra in Section~\ref{results}, in which we will study
the production of heavy neutral Higgs bosons and their different
decay modes.
In summary, in our exploration of the MSSM ways, we assume
that the LHC will not discover any colored superpartner but weakly-interacting ones
(neutralinos and/or charginos).

\subsection{Plausible routes from high-scale theories to Slim SUSY}
\label{plausibility}

In this section we introduce arguments of plausibility in order to inspire this kind of low-energy spectra from general high-scale theories of SUSY breaking. This should be understood as a qualitative discussion, as we are not building a specific model, because we prefer to work in a general setting. It is also important to clarify in this sense that we are not generating the SUSY spectra at low energies from the high-energy scale through renormalization group evolution.

Such a class of mass spectra might emerge from different theoretical realizations of SUSY breaking, including
PeV-scale supersymmetry~\cite{Wells:2004di} and pure gravity mediation~\cite{Ibe:2011aa,Ibe:2012hu,Bhattacherjee:2012ed}.
The main idea behind them is to give rise to the masses of the  supersymmetric particles
through dynamical breaking of supersymmetry, where the chiral supermultiplet $S$, which breaks SUSY, is 
charged under some symmetry.
Following~\cite{Wells:2004di} and~\cite{Ibe:2011aa}, this superfield $S$ is parametrized by
\begin{equation}
S = S + \sqrt{2} \psi \theta + F_S \theta^2 \,,
\end{equation}
whose nonzero $F_S$ component is the source of supersymmetry breaking.
The scalar masses are generated at tree-level by
\begin{equation}\label{scalar_masses}
\int d^2\theta d^2\bar\theta \, c_ i \,\frac{S^\dag S}{M_\text{Pl}^2} \Phi_i^\dag \Phi_i \to c_ i\frac{F_S^\dag F_S}{M_\text{Pl}^2} \phi_i^\ast \phi_i \,,
\end{equation}
where $M_\text{Pl}$ is the reduced Planck scale and
$c_i$ ($i = H, Q, U, D, L, E$) are in principle coefficients of ${\cal O}$(1).
Therefore, one obtains $m_0 \sim c_i \, m_{3/2}$ with $m_{3/2}^2 = \langle F_S^\dag F_S \rangle / M_\text{Pl}^2$.
On the other hand, gaugino masses would arise from the anomaly mediation and read as
\begin{equation}
M_{\lambda_a} = \frac{\beta(g_a)}{g_a} m_{3/2} \,,
\end{equation}
where the beta function is given at one-loop by $\beta(g_a) = b_a g_a^3/(16 \pi^2)$ and $b_a$ denotes the coefficients of renormalization-group equations (RGEs) of $g_a$.

Thus, we shall study the constraints on the MSSM parameters in order to have $m_{h^0} \simeq$ 125 GeV,
and the predictions for masses and couplings of the heavy Higgs states ($H^0$, $A^0$).
In principle, their mass could be as low as the LHC admits, i.e. $m_{H^0}$, $m_{A^0} =$ 200$-$600 GeV, 
which are much lighter than the sfermion masses. In order to obtain this hierarchy in the
simplest way, for gravitino masses
of order 10 TeV, we can simply assume that $c_{H_u} \simeq c_{H_d} = {\cal O}(10^{-1})$ and 
$c_{Q_{1,2}} \simeq c_{U_{1,2}} \simeq c_{D_{1,2}} = {\cal O}(10)$. For the third generation of sfermions, we shall take
$c_{Q_3} \simeq c_{U_3} \simeq c_{D_3} = {\cal O}(1)$.

This pattern could be explained for instance in a supersymmetric theory of flavor based
on the Froggatt-Nielsen mechanism~\cite{Froggatt:1978nt},
which would be invoked not only to generate the Yukawa couplings, but also to 
explain the difference between matter and Higgs superfields. Namely, when one considers 
flavor symmetries,  the matter supermultiplets containing  the quarks, leptons, and their superpartners,
are usually  charged under a flavor symmetry, while the Higgs multiplets are assigned
as singlets.  Thus, they would have different behavior when the flavor symmetry is broken.
Along this line, we could follow~\cite{DiazCruz:2005qz}, where it is proposed to use
a SUSY-breaking sector which generates the CP violating phases of the MSSM. This model uses a $U(2)_H$ flavor symmetry, and the resulting SUSY-breaking pattern is such that sfermions of first and second generation could receive contributions from one source to the soft-breaking masses, while the sfermions of third family and the Higgs doublets could get their soft-masses
from another source. This is precisely the pattern of soft-breaking masses
that we are advocating in this paper.

There are other possibilities that one could imagine, such as the 
heterotic string constructions~\cite{heterotic-strings}, where the Higgs and the third 
family arise in the untwisted sector, while the first and second families belong to the twisted sector,
resulting in a UV realization of Natural SUSY~\cite{Badziak:2012yg}.
It is also possible that the Higgs multiplets correspond to pseudo-Goldstone bosons of a global
symmetry~\cite{pseudoGoldstone} or even they could be composite states~\cite{compositeHiggs} 
and they would not have consequently the same mass as the sfermions.

\section{Higgs mass and LHC constraints}
 \label{constraints}

The first constraint that any SUSY model should fulfill nowadays is the occurrence of a light
Higgs state with a mass near $m_{h^0}\simeq 125$ GeV.
After considering the results, including  statistical and systematic uncertainties reported by
ATLAS~\cite{:2012gk} and CMS~\cite{:2012gu}, we consider a central value for $m_{h^0}$
of 125 GeV and an uncertainty of $\pm 3$ GeV, i.e. we accept a value of $m_{h^0}$
in our numerical analysis if it lies within the range [122 GeV, 128 GeV].

\begin{figure}[t!]
\begin{center}
\begin{tabular}{cc}
\includegraphics[width=75mm]{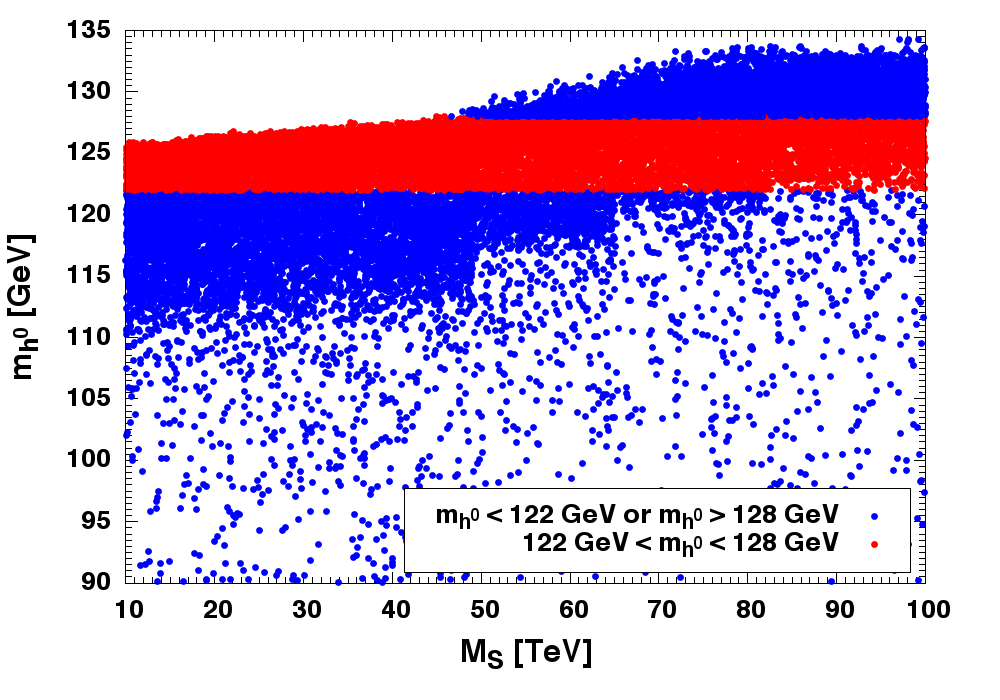} &
\includegraphics[width=75mm]{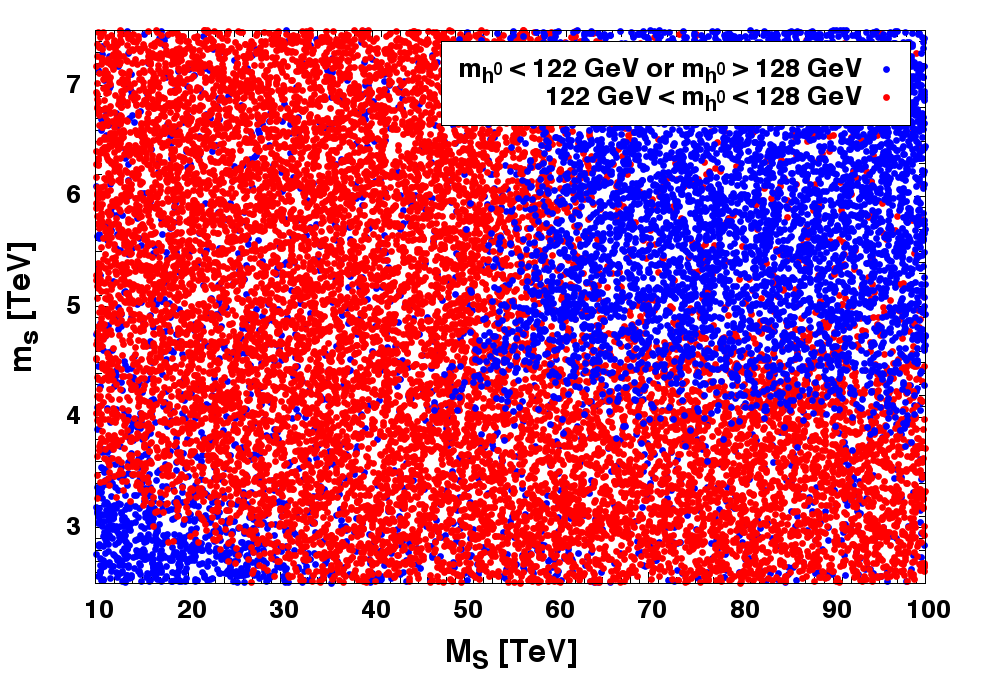}
\end{tabular}
\caption{Scatter plots of the allowed regions in parameter space for $m_{h^0}$.
Left panel: $m_{h^0}$ as a function of $M_S$.
Right panel: $m_{h^0}$ in the plane $m_s$$-$$M_S$.
In both plots, red dots are for 122 GeV $< m_{h^0} <$ 128 GeV and blue dots represent values of $m_{h^0}$
smaller than 122 GeV or larger than 128 GeV. Values for the rest of the parameters were varied randomly,
with $A_t =$ 0.}\label{fig:mh0}
\end{center}
\end{figure}

For this, we have generated scatter plots (by means of the use of the code
{\tt SuSpect}~\cite{Djouadi:2002ze}) included in Figure~\ref{fig:mh0} that show the different regions in
parameter space where $m_{h^0}$ lies between 122 GeV and 128 GeV (red dots)
or falls outside this range (blue dots). On the left panel we can see the behavior
of $m_{h^0}$ with $M_S$. The dependence on this parameter is not so pronounced as
on $m_s$ (see our previous work~\cite{Arganda:2012qp}) but it is not negligible, since
we are not decoupling the sfermions of the first and second generations. It is clear that it is
possible to obtain values of $m_{h^0}$ close to its current experimental value for
all the values of $M_S$ considered here, within the range [10 TeV, 100 TeV].
On the right panel of Figure~\ref{fig:mh0} the behavior of $m_{h^0}$ with $M_S$
and $m_s$ is displayed. For low values of $M_S$, close to 10 TeV, values of $m_s$
smaller than 3.5 TeV do not allow to get values of $m_{h^0}$ in the valid range. As $M_S$ increases,
the range of $m_s$ that generates correct values of $m_{h^0}$ become larger. For values of $M_S$
between 30 and 50 TeV, stop masses in the range [2.5 TeV, 7.5 TeV] drive us to 122 GeV $< m_{h^0} <$ 128 GeV.
From $M_S \simeq$ 50 TeV, this window starts to close and only low values of $m_s$, between
2.5 TeV and 4 TeV, result in proper values of $m_{h^0}$. We can conclude from these two plots that both parameters
$m_s$ and $M_S$ are important in order to obtain correct values of $m_{h^0}$, although
the dependence on the former is stronger.

\begin{figure}[t!]
\begin{center}
\begin{tabular}{c}
\includegraphics[width=100mm]{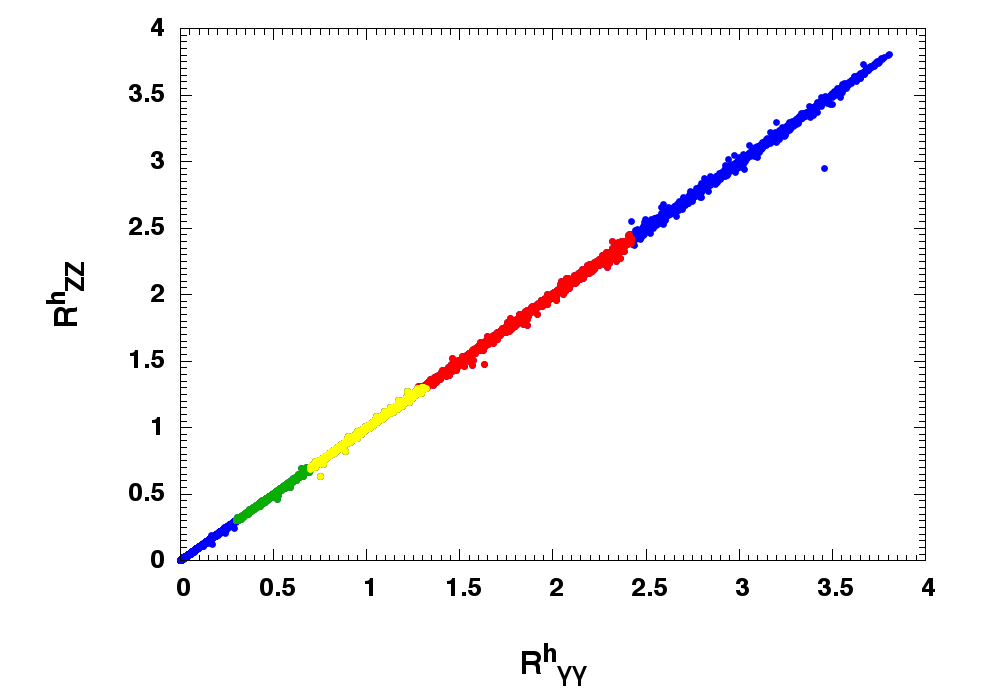}
\end{tabular}
\caption{Correlation between $\gamma \gamma$ and $ZZ$ 
signal rates for the light Higgs boson $h^0$.
Red and green dots are for 0.7 $< R_{\gamma \gamma}^h <$ 2.42 (95\% C.L.) and
0.3 $< R_{ZZ}^h <$ 1.3 (95\% C.L.), respectively; yellow dots represent points of the parameter space
that fulfill both previous requirements; blue dots do not satisfy any of them.}\label{fig:Rh0}
\end{center}
\end{figure}

The next constraint that needs to be satisfied is the strength of the
SM-like Higgs signal observed at the LHC~\cite{ATLAS-CMS_Hsignals}. Namely, in order to reproduce 
the signal rate for the  SM-like Higgs signals with $m_{h^0}\simeq 125$ GeV, within Slim SUSY scenarios,
we show in Figure~\ref{fig:Rh0} the ratios defined as follows:
\begin{equation}\label{eq:Rs}
 R^{h,H}_{XX} = \frac{\sigma(gg\to h^0, H^0)}{\sigma(gg\to h_\text{SM})} \, \frac{\text{BR}(h^0, H^0 \to XX)}{\text{BR}(h_\text{SM} \to XX)} 
\end{equation}
for $X=\gamma, \, Z$ (for the calculation of these ratios we have used the code
{\tt FeynHiggs}~\cite{FeynHiggs}).
In these plots, red and green dots are for 0.7 $< R_{\gamma \gamma}^h <$ 2.42 (95\% C.L.) and
0.3 $< R_{ZZ}^h <$ 1.3 (95\% C.L.), respectively,  while yellow dots represent points of the parameter space
that fulfill both previous requirements and blue dots do not satisfy any of them.
This figure shows that plenty of points satisfy both constraints for $h^0$ within the Slim SUSY scenario.

\begin{figure}[t!]
\begin{center}
\begin{tabular}{c}
\includegraphics[width=100mm]{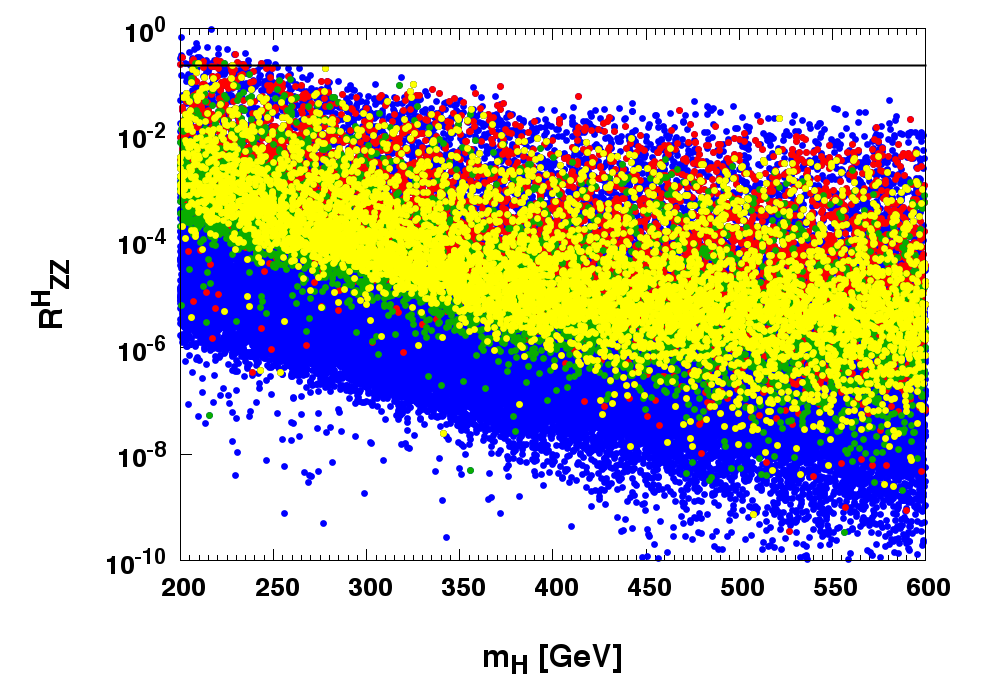}
\end{tabular}
\caption{$R_{ZZ}^H$ as a function of the heavy Higgs mass $m_{H^0}$.
Red and green dots are for 0.7 $< R_{\gamma \gamma}^h <$ 2.42 (95\% C.L.) and
0.3 $< R_{ZZ}^h <$ 1.3 (95\% C.L.), respectively; yellow dots represent points of the parameter space
that fulfill both previous requirements; blue dots do not satisfy any of them.}\label{fig:RHZZ}
\end{center}
\end{figure}

On the other hand, in Figure~\ref{fig:RHZZ} we have displayed the values of the 
corresponding quantity $R^H_{ZZ}$ for the heavy CP-even Higgs boson $H^0$, which can also be constrained
from current LHC searches. This discussion is only based on the decay mode $H^0 \to ZZ^*$, 
while the results from the other relevant decays of $H^0$ are left for the following section.
The ratio $R^H_{ZZ}$ is also defined in Eq.~(\ref{eq:Rs}), and it is presented as
a function of the heavy Higgs mass, for those points where $R^h_{XX}$ lay in the ranges defined
in Figure~\ref{fig:Rh0}. For illustration, we also display in this figure the value $R^H_{ZZ}=0.2$,
which is the minimum value that LHC has excluded for the mass range 200$-$600 GeV~\cite{ATLAS-CMS_HZZ},
which is well above the values obtained for $H^0$ within the Slim SUSY scenario.

\section{Decays and production of heavy neutral Higgs bosons at the LHC}
\label{results}

\begin{figure}[t!]
\begin{center}
\begin{tabular}{cc}
\includegraphics[width=75mm]{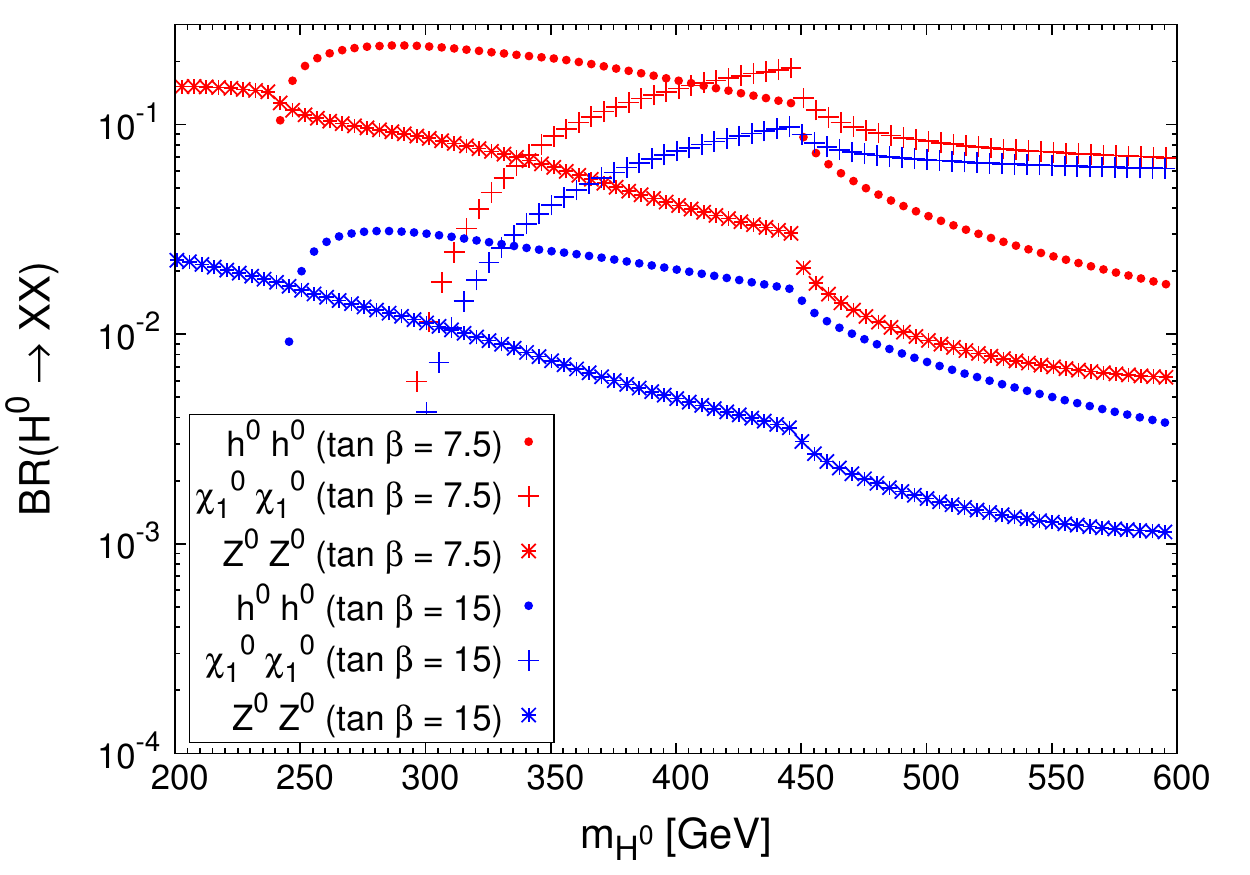} &
\includegraphics[width=75mm]{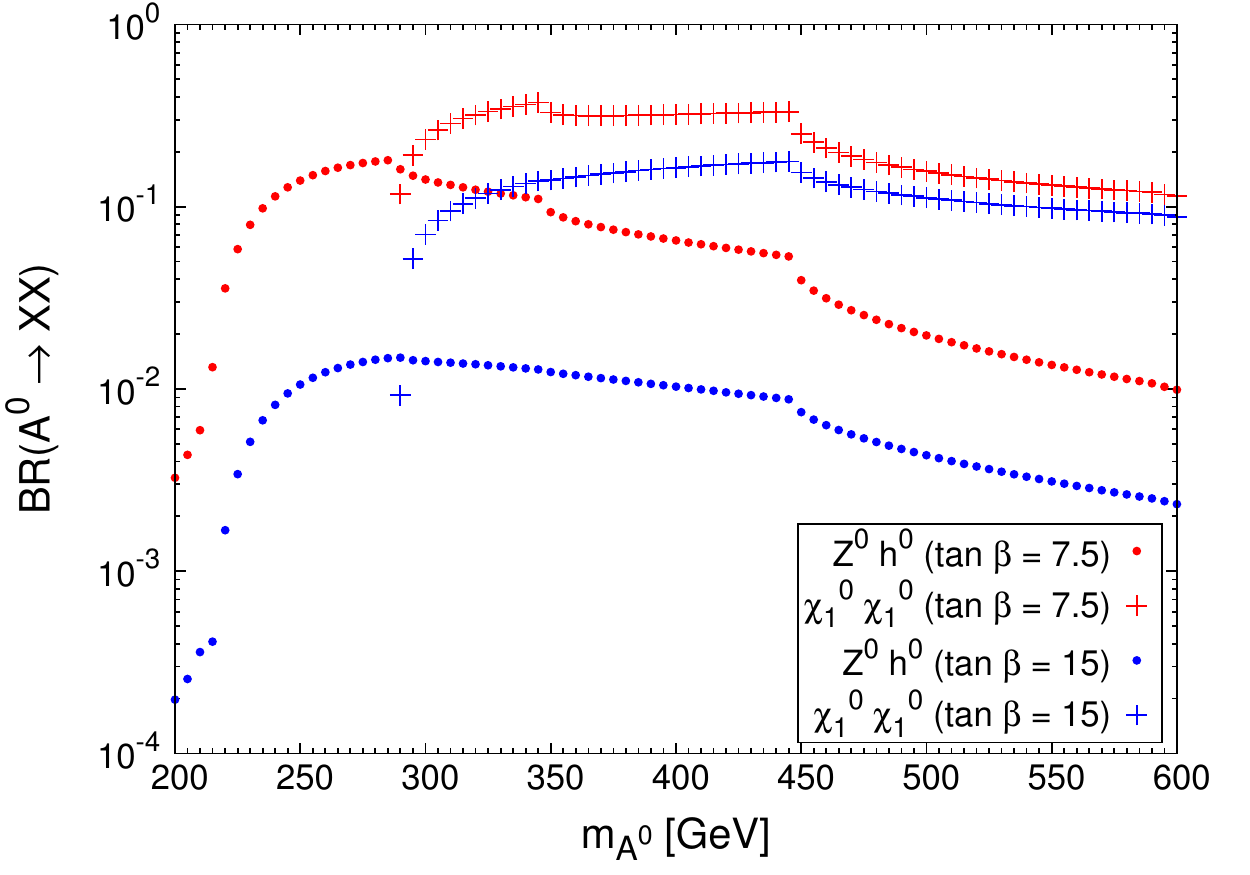}
\end{tabular}
\caption{$H^0$ (left panel) and $A^0$ (right panel) decay channels as a function of $m_{H^0}$
and $m_{A^0}$, respectively, for $M_S =$ 35 TeV, $m_s =$ 7.5 TeV, $A_t =$ 0,
$M_1 =$ 150 GeV, $M_2 =$ 1 TeV, $M_3 =$ 3 TeV, $\mu =$ 300 GeV and $\tan\beta =$ 7.5
(in red) or $\tan\beta =$ 15 (in blue).
}\label{fig:BRs}
\end{center}
\end{figure}

Given that our samples satisfy the constraints on the SM-like Higgs signal at the LHC, we would like now
to identify new signals of the heavy Higgs states, which could be searched at the LHC.
We know from~\cite{Arganda:2012qp} that for the most of the regions
of the parameter space, the dominant decay modes are $H^0 \to b \bar b, \tau^+ \tau^-$
and $A^0 \to b \bar b, \tau^+ \tau^-$ for $\tan\beta \gtrsim$ 10, or
$(H^0, A^0) \to t \bar t$, if it is kinematically allowed, for low values of $\tan\beta$.
However, the corresponding signatures of these decay channels are very difficult to distinguish
from the SM background.

Therefore, we show in Figure~\ref{fig:BRs} the results for
the branching ratios of the most relevant decays of $H^0$ and $A^0$ Higgs bosons
which could shed light on some new physics.
On the left panel we see the dependence of BR($H^0\to h^0 h^0$), BR($H^0\to Z^0 Z^0$)
and BR($H^0\to \tilde \chi_1^0 \tilde \chi_1^0$) on $m_{H^0}$.
On the one hand, for low values of $\tan\beta$ (points in red)
both decay modes $H^0\to h^0 h^0$ and $H^0\to \tilde \chi_1^0 \tilde \chi_1^0$
can obtain important branching ratios (BR($H^0\to h^0 h^0$) $\simeq$ 0.2 for $m_{H^0} \simeq$
250$-$300 GeV and BR($H^0\to \tilde \chi_1^0 \tilde \chi_1^0$) $\simeq$ 0.2 for $m_{H^0} \simeq$ 450 GeV).
On the other hand, if we double the value of $\tan\beta$ (points in blue), these branching ratios
decrease drastically (around one order of magnitude for $H^0\to h^0 h^0$ and softer for the invisible decay),
because of the enhancement proportional to
$\tan\beta$ on $b \bar b$ and $\tau^+ \tau^-$ decay modes.
It is important to note that the large branching ratios of the $H^0$ invisible decay,
compared to our results obtained in~\cite{Arganda:2012qp}, are due to the choice of
input parameters. Concretely, the values of $M_1$ and $\mu$ chosen in Figure~\ref{fig:BRs}
produce a large gaugino-higgsino mixing, necessary to have non-negligible Higgs-neutralino-neutralino
couplings.
A similar behavior is depicted on the right panel of Figure~\ref{fig:BRs} for
$A^0 \to Z^0 h^0$ and $A^0 \to \tilde \chi_1^0 \tilde \chi_1^0$ decay channels,
as a function of $m_{A^0}$.
In this case, we can obtain values of BR($A^0 \to \tilde \chi_1^0 \tilde \chi_1^0$)
even larger (up to 0.4 for $m_{A^0} \simeq$ 350 GeV).
It is also remarkable that even for $\tan\beta =$ 15, the branching ratio of this invisible decay
is always around 0.1. The decay mode $A^0 \to Z^0 h^0$ is also interesting and
can reach a branching ratio of 0.2 for $\tan\beta \simeq$ 7.5 and $m_{A^0} \simeq$ 290 GeV.
However, it is also very sensitive to $\tan\beta$, as $H^0\to h^0 h^0$ channel, and for $\tan\beta \simeq$ 15
suffers a large suppression, around one order of magnitude or more.
To sum up, we notice from these two plots that the decay modes $H^0\to h^0 h^0$, $A^0 \to Z^0 h^0$,
as well as the invisible decays into the  LSP neutralinos $(H^0,A^0) \to \tilde \chi_1^0 \tilde \chi_1^0$,
achieve measurable branching ratios that could be interesting to further study.

\begin{figure}[t!]
\begin{center}
\begin{tabular}{c}
\includegraphics[width=100mm]{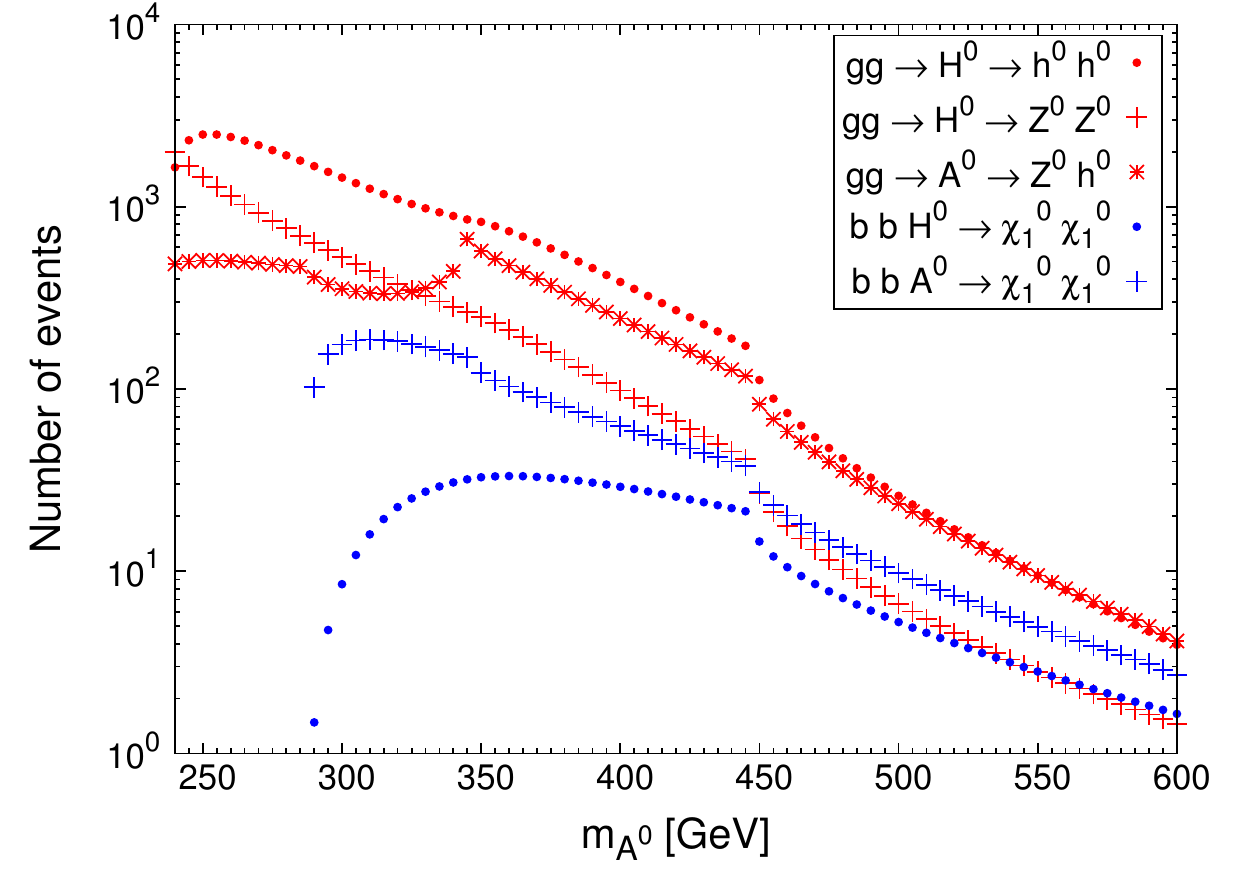}
\end{tabular}
\caption{Number of events expected at the LHC for $H^0 \to h^0 h^0$, $Z^0 Z^0$
and $A^0 \to Z^0 h^0$ through Higgs production via gluon fusion (in red),
and ($H^0$, $A^0$) $\to \tilde \chi_1^0 \tilde \chi_1^0$
through Higgs production associated with a pair of bottom quarks (in blue),
for a total integrated luminosity of ${\cal L} =$ 23 fb$^{-1}$ and
a center-of-mass energy of $\sqrt{s} =$ 8 TeV.
The input parameter are chosen as in Figure~\ref{fig:BRs}:
$M_S =$ 35 TeV, $m_s =$ 7.5 TeV, $A_t =$ 0,
$M_1 =$ 150 GeV, $M_2 =$ 1 TeV, $M_3 =$ 3 TeV, $\mu =$ 300 GeV and $\tan\beta =$ 7.5.
}\label{fig:Nevents}
\end{center}
\end{figure}

We show next, in Figure~\ref{fig:Nevents}, the expected number of events for these signals at the LHC,
calculated as
\begin{equation}
N_\text{events} = \sigma(H^0, A^0) \times \text{BR}(H^0, A^0 \to XX) \times {\cal L} \,,
\end{equation}
where $\sigma(H^0, A^0)$ is the production cross section of $H^0$ and $A^0$
(computed with {\tt FeynHiggs} too), respectively,
and ${\cal L}$ is the total integrated luminosity of the LHC.
We can see from this plot that the most promising process,
in order to obtain measurable new physics signals, is the production of
$H^0$ via gluon fusion and the consequent decay into two light Higgs bosons,
with more than $2 \times 10^3$ expected events for the current ${\cal L} =$ 23 fb$^{-1}$
and low values of $m_{A^0}$. The production of the pseudoscalar Higgs boson $A^0$
via gluon fusion and its decay into $Z^0 h^0$ is also an interesting process,
but the number of events expected is lower, $1 \times 10^3$ at the most
for $m_{A^0} \simeq$ 350 GeV. Both processes are not sufficient to
distinguish a 2HDM from Slim SUSY scenarios and we have to
resort to the invisible decays of $H^0$ and $A^0$. The problem in this case is that
we need some particles in the final state to be tagged in order to identify the
missing transverse energy produced by the two LSP neutralinos.
Thus, for the processes with neutralinos in the final state,
we consider the production of $H^0$ and $A^0$ associated with
a pair of bottom quarks, which have to be tagged \cite{ourhbblhc}.
The number of events predicted in these latter processes are even much lower,
less than 70 for $H^0$ with $m_{A^0}$ around 350 GeV and close to 200
for $A^0$ with $m_{A^0} \simeq$ 300 GeV.
Moreover, these numbers will be reduced after $b$-tagging process.
Nevertheless, the combination of the production of $H^0$, via gluon fusion,
decaying into $h^0 h^0$ and the production of $A^0$, associated with a pair of bottom quarks,
decaying into two LSP neutralinos could provide a clear hint of the presented Slim SUSY scenarios.

\section{Conclusions}
\label{conclusions}

The recent LHC results on the mass of the new SM-like Higgs boson,
$m_{h_\text{SM}} \simeq$ 125 GeV, as well as the ${\cal O}$(TeV) direct bounds on the mass 
of colored superpartners, suggest that a heavy SUSY scale should be part of the surviving
MSSM. In this paper we have proposed an alternative MSSM scenario, called Slim SUSY, 
which has gluinos and sfermions with multi-TeV masses.
Gluinos and sfermions of third generation have masses of ${\cal O}$(TeV), 
in order to account for the Higgs mass value ($m_{h_\text{SM}} \simeq$ 125 GeV),
while sfermions of the first and second generations are assumed to be heavy
enough (${\cal O}$(50$-$100 TeV)) to solve the SUSY and CP flavor problems, or at least to ameliorate them.
The Slim SUSY scenario contains gauginos/higgsinos near the EW scale; 
in this regard, it is similar to some MSSM scenarios proposed in the 
literature, such as Natural SUSY, pure gravity mediation, Split and Spread SUSY, among others.
However, the scenario includes, as a new feature,  the heavy MSSM Higgs bosons ($H^0$, $A^0$, $H^\pm$) 
near the EW scale too.  
In fact, these Higgs scalars could be searched at the LHC and provide the first 
signature  of SUSY at the EW scale, together with a DM candidate. In summary, within our 
exploration of the possible ways that SUSY could be realized in nature, we are assuming that
no strongly- but only weakly-interacting superpartners will be discovered at the LHC.

We have discussed the theoretical constraints on Slim SUSY and have found regions of parameters
where the light Higgs boson $h^0$ lays within the mass range [122 GeV, 128 GeV], and its couplings satisfy
LHC constraints too. 
We have also imposed the constraints from LHC Higgs searches through the $ZZ^*$ channel  
for the heavy CP-even Higgs boson $H^0$, finding that most of the points generated satisfy this bound.
Then, we have identified distinctive heavy Higgs signals 
that could be searched at the LHC, including the decay modes
$H^0\to h^0 h^0$ and $A^0 \to Z^0 h^0$, as well as the invisible decays into the 
LSP neutralinos $(H^0, A^0) \to \tilde \chi_1^0 \tilde \chi_1^0$.

The mood of the 90's was to expect that LEP would start the detection of the full spectrum 
of superpartners of the MSSM, and the task would be completed at the LHC. We have learned by now that 
the possible realization of SUSY in nature, and its detection at the LHC, will not be as exuberant 
as it was thought then, but rather slim.

\section*{Acknowledgments}

The authors are indebted to Mart\'{\i}n Tripiana for providing the tools to plot Figure~\ref{fig:mass_spectrum}.
E.A. is financially supported by a MICINN postdoctoral
fellowship (Spain), under grant No. FI-2010-0041, and thanks IFLP-CONICET for hospitality and support.
J.L.D.C. acknowledges support from VIEP-BUAP and CONACYT-SNI (Mexico) and hospitality at IFLP.
This work has been partially supported by ANPCyT (Argentina) under
grant No. PICT-PRH 2009-0054 and by CONICET (Argentina) PIP-2011 (E.A., A.S.).

\bibliographystyle{unsrt}

\end{document}